# *Topological magnetic dipoles and emergent field bundles in a ferromagnetic microstructure by X-ray magnetic vector tomography*

Javier Hermosa[1,2], Aurelio Hierro-Rodríguez[1,2], Carlos Quirós[1,2], José I. Martín[1,2], Andrea Sorrentino[3], Lucía Aballe[3], Eva Pereiro[3], María Vélez[1,2] and Salvador Ferrer[3]

[1]*Depto. Física, Universidad de Oviedo, 33007 Oviedo, Spain.*
[2]*CINN (CSIC – Universidad de Oviedo), 33940 El Entrego, Spain.*
[3]*ALBA Synchrotron, 08290 Cerdanyola del Vallès, Spain.*

**Abstract.** Advanced vector imaging techniques provide us with 3D maps of magnetization fields in which topological concepts can be directly applied to describe real-space experimental textures in non-ideal geometries. Here, the 3D magnetization of a low symmetry permalloy microstructure is obtained by X-ray vector magnetic tomography and analysed in detail in terms of topological charges and emergent fields. A central asymmetric domain wall with a complex 3D structure is observed in which magnetization chirality transitions are mediated by Bloch points arranged in several dipoles and a triplet. The ideal spherical symmetry of the emergent field of an isolated monopole is severely modified due to shape effects of the permalloy microstructure. Emergent field lines aggregate into bundles that either connect adjacent Bloch points within a topological dipole or tend towards the surface. These bundles may present different textures such as helical vortices or half merons, depending on asymmetries and confinement, but are constrained by topological charge conservation in a given sample volume. This precise description of the singularities in realistic systems, enabled by the quantitative experimental information on magnetization vector fields, can significantly improve our understanding of topological constraints in 3D magnetic systems and provide advancements in the design of magnetic devices.

# I. INTRODUCTION

Topological textures in ferromagnetic systems such as vortices, skyrmions, Bloch points or hopfions [1-3] are the focus of an intense research effort as possible information carriers in novel memory and logic devices [1,4]. Bloch points, i. e., singular points in the magnetization, appear in many different systems with 3D magnetization configurations such as stripe domain patterns of multilayers [5,6], skyrmion lattices [7,8] or perpendicular magnetic anisotropy ferrimagnets [9]. The physics of these topological textures is key for both dynamic and quasi-static processes. Bloch points drive the in-plane magnetisation reversal processes in stripe domain systems following strong selection rules based on the topological characteristics of bifurcations [6,10]. Also, in the highly confined nanowire geometry, the presence of Bloch points and surface textures is essential to understand domain wall propagation depending on chirality and polarity [1, 11-15].

Analytical calculations and micromagnetic simulations have been widely used to study magnetic textures [2,16] but magnetic singularities remain a challenging multiscale problem [17] particularly in the presence of magnetostatic confinement [18,19]. From an experimental point of view, the development of magnetic vector tomography [20-22] and laminography [23] has allowed for the quantitative characterization of magnetization vector fields in thin films and nanostructures. Experimental 3D magnetization maps allow moving beyond a qualitative description of magnetization at domain walls and magnetic textures [11,24-28] and to use the full power of vector analysis methods for the description of magnetic singularities in terms of topological charges and emergent fields [22,29-31]. For instance, topological monopoles and dipoles have been recently identified in ferromagnetic and ferrimagnetic systems with perpendicular anisotropy [8,21] and closed vortex rings decorated by magnetic singularities have been observed in microstructures with fully closed magnetization [30].

In this work, 3D magnetization textures in the magnetization vector field $\boldsymbol{m}(\boldsymbol{r})$ of a low-symmetry ferromagnetic microstructure have been studied by magnetic X-ray vector tomography. Magnetostatic confinement breaks the spherical symmetry of ideal monopoles creating strong distortions in the emergent fields that surround the observed singularities. Bloch point dipoles and triplets appear connected by bundles of emergent field lines carrying a fractional emergent field flux. These bundles, which can be considered as generalized low symmetry vortices, determine the 3D domain wall configuration in the microstructure, constrained by topological charge conservation within a given closed surface.

## II. EXPERIMENTAL RESULTS

A 140 nm thick permalloy boomerang-shaped structure was fabricated by e-beam lithography, DC magnetron sputtering and lift-off [32] on a $Si_3N_4$ membrane (Ted Pella, 21501-10) with 50 nm thickness and 750 $\mu$m × 750 $\mu$m size. Sample thickness was chosen well above 50 - 60 nm, the thin film limit of symmetric Néel walls in extended permalloy films [16] and of transverse/vortex walls in permalloy nanostrips [33], so that the competition between exchange, anisotropy and magnetostatics can give rise to domain walls with a complex internal structure [33-38] as e. g. in Landau [33], asymmetric Néel or Bloch [16] walls. We chose an elongated low-symmetry boomerang shape, as shown Fig. 1, in order to favour a closed flux domain structure with a long central wall. The sample was studied at remanence after saturating it with a 3000 Oe in-plane field along its symmetry axis ($y$-axis in Fig. 1). This field preparation favours the independent nucleation of reverse domains at different points in the boomerang and, hence, the observation of transitions between different equivalent domain wall configurations in the final remanent state.

The 3D magnetization vector field $\boldsymbol{m}(\boldsymbol{r})$ was measured by X-ray vector magnetic tomography (XVMT) at the MISTRAL beam line (ALBA Synchrotron) [39-41] making use of the angular dependent contrast in Magnetic Transmission X-ray Microscopy (MTXM). Figure 1 shows several MTXM projection images, measured at the Fe $L_3$ absorption edge (706.8 eV) with different angles of incidence (angle geometry as sketched in Fig. 1(a)). MTXM contrast is associated with the projection direction of the magnetization along the X-ray beam direction so that for $\phi = 0^0$ the projections are sensitive to $m_x$ and $m_z$ whereas for $\phi = 90^0$ contrast is given by $m_y$ and $m_z$ (see Figs. 1(b-c)) [6]. Normal incidence projections ($\theta = 0^0$), which are sensitive only to the out-of-plane magnetization component $m_z$, display two strong bright/dark lines running parallel along the microstructure centre. They indicate a central DW that separates magnetic domains with in-plane magnetization.

The combined information of two tilt-series of projections acquired at $\phi = 0^0$ and $\phi = 90^0$ with variable $\theta$ is used to reconstruct the full 3D magnetization configuration of the microstructure with a magnetic vector tomography algorithm [21,40] (see details in the Supplementary information [41]). A central DW with a non-trivial 3D configuration is observed in the magnetic tomogram separating in-plane magnetized domains parallel to the microstructure edge (see $m_x$ and $m_y$ maps at the microstructure middle plane $z = 0$ in Figs. 1(d-e)) except for a diamond transverse domain near the right end of the boomerang. Thus, the natural reference system to describe the magnetic configuration in this curved geometry will be a right-handed set of unit vectors defined locally by the central line of the microstructure (see sketch in Fig. 1(e)): parallel ($\boldsymbol{u}_\parallel$) and transverse ($\boldsymbol{u}_\perp$) within the sample plane and out-of-plane ($\boldsymbol{u}_z = \boldsymbol{u}_\parallel \times \boldsymbol{u}_\perp$). The central DW in the microstructure is defined by the $m_\parallel = 0$ boundary between $\pm m_\parallel$ domains (see solid yellow line in Fig. 1(e)).

The strongest out-of-plane magnetization signal appears close to the microstructure centre (see $m_z$ map at $z = 0$ in Fig. 1(f)), but it is quite different from a simple unipolar Bloch DW core: it contains both up/down $m_z$ regions (red/blue stripes in Fig. 1(f)) of different width and intensity depending on sample location. This configuration is typical of asymmetric Bloch and Néel walls in films of intermediate thickness [16] and is related with flux-closure effects across the sample thickness. The consequence is that an additional $m_z = 0$ boundary appears. It runs almost parallel to the central $m_\parallel = 0$ DW between in-plane domains but with a non-trivial configuration marked by kinks and meanders in which $m_z$ contrast is reversed (see central white line that marks the transition from red ($+m_z$) to blue ($-m_z$) contrast in Fig. 1(f)). Thus, it is an ideal configuration to search for magnetic singularities such as Bloch points, which are characterized by the condition of zero magnetization modulus [16,42], $|\boldsymbol{m}| = \sqrt{m_x^2 + m_y^2 + m_z^2} = 0$.

### III. RESULTS AND DISCUSSION
*A. Bloch Points: Magnetization Texture and Topological Charge.*

Bloch points were searched through the microstructure at the crossing points of $m_x = 0$, $m_y = 0$, and $m_z = 0$ isosurfaces [18] with the results shown in Fig. 2. Most of them appear within 20-30 nm of the microstructure central plane $z = 0$ (see Table 1), decorating the kinks and turns of the $m_z = 0$ boundary at the DW core (see dashed arrows in Fig. 2(a)). For example, Bloch point B1 is sitting at a DW kink near the left end of the microstructure and is composed of a head-to-head (H2H) line wall surrounded by a magnetization vortex in an almost perpendicular plane (see Fig. 2(b)). Next to it, in a second DW kink, we find Bloch point B2 also in a circulating configuration but with a tail-to-tail (T2T) central line wall (see Fig. 2(c)). As shown in Figs. 2(b-h), all the Bloch points in the microstructure present a circulating magnetization configuration which, according to analytical predictions, should be the preferred one over radial hedgehogs in extended systems [17].

In an initial geometrical characterization, each Bloch point is defined by its core polarity reversal (H2H or T2T), spatial core orientation (defined by $t$, a vector tangent to the central line wall with a positive projection along $u_\parallel$ as shown in the sketches of Fig. 2) and circulation sense (defined as clockwise (CW) or counter-clockwise (CCW) relative to the positive sense of $t$), as indicated in Table 1. We observe a trend of H2H and T2T Bloch points to appear in pairs with the same circulation sense (e. g. B1-B2, B4-B5 or B6-B7), with the exception of B3 (H2H and CCW) sitting next to B4 (H2H and CW).

A more fundamental representation can be obtained from the 3D magnetization $m(r)$ in terms of emergent fields $\mathbf{B^e}$ and topological charges. Briefly, the emergent field $\mathbf{B^e}$ describes the effect of magnetic vorticity on the Hamiltonian of the ferromagnet [43] and is given at each sample point by [44,45]

$$B_i^e = \frac{\hbar}{2} \epsilon_{ijk}\, \boldsymbol{m} \cdot \partial_j \boldsymbol{m} \times \partial_k \boldsymbol{m}, \qquad (1)$$

where $\epsilon_{ijk}$ is the Levi-Civita tensor. With this definition $\mathbf{B^e}$ is divergence-free as long as the condition $|\boldsymbol{m}| = 1$ is met and, correspondingly, $\oiint \mathbf{B^e} \cdot d\boldsymbol{S} = 0$ for any closed surface [46]. The number of singular points with $|\boldsymbol{m}| = 0$ is calculated from the flux of $\mathbf{B^e}$ through any closed surface enclosing them as [29,44]

$$Q = \frac{1}{4\pi\hbar} \oiint \mathbf{B^e} \cdot d\boldsymbol{S} \qquad (2).$$

$Q$ is the topological charge and, in analogy with the Gauss theorem of electrostatics [45], the singularities in the magnetization become topological monopoles with $Q = +1$ or $Q = -1$ depending on whether they are sources or sinks for $\mathbf{B^e}$ lines. Thus, eq. (2) implies a direct link between the magnetic configuration at any closed surface (where the flux of $\mathbf{B^e}$ is calculated) and the number of singularities inside it [29]. In an ideal system the emergent field created by a monopole is simply $\mathbf{B^e} = \frac{Q}{4\pi r^3} \boldsymbol{r}$ with a Coulomb-like spherical configuration [46] but theoretical models suggest significant distortions in confined geometries [17,19].

The last column in Table 1 displays the experimental topological charges of each Bloch point calculated as the flux of $\mathbf{B^e}$ across a cubic box of lateral size $L = 70$ nm containing the singularity. $Q$ alternates along the central DW between +1 and -1 for T2T and H2H Bloch points, respectively. Several emergent field dipoles are identified composed of pairs of opposite topological monopoles such as B1-B2 or B6-B7 joined by a short wall segment with strong $+m_z$ component. The closely spaced group of three Bloch points near the boomerang vertex is composed of two negative monopoles (B3-B4) joined to a positive monopole (B5) by a zig-zag $-m_z$ segment. The correlation of these topological dipoles and triplets with the surrounding magnetization textures will be analysed in detail in the following.

### *B. Bloch Point Dipole and Emergent Field Bundles.*

The emergent field configuration near Bloch points B6-B7 illustrates the mechanism for DW chirality reversal by a polarity change. In the close-up view of Bloch point B6 (Fig. 3(a)), emergent field lines look approximately radial. However, on a larger length scale containing the complete topological dipole (Fig. 3(b)), it is clear that they deviate from this ideal configuration and present a high density of $\mathbf{B^e}$ lines going from $Q^{B7} = +1$ to $Q^{B6} = -1$. These distortions appear also in the plot of $\mathbf{B^e} \cdot d\mathbf{S}$, the density of emergent flux, on a 50 nm diameter spherical shell surrounding B6 (see Fig. 3(c)): there is a blue band in the $\boldsymbol{u_\parallel}$ - $\boldsymbol{u_z}$ section with stronger dark blue spots marking the crossing of bundles of $\mathbf{B^e}$ lines and a yellow region around the $\boldsymbol{u_\perp}$ pole due to negligible flux of $\mathbf{B^e}$ in the transverse direction. These bundles of $\mathbf{B^e}$ lines are equivalent to the high magnetic vorticity tubes observed in soft GdCo microstructures joining different topological defects [30]. However, in the present case, magnetostatic confinement by the microstructure surfaces splits the emergent field lines entering into B6 into four bundles: two along $\boldsymbol{u_\parallel}$, i. e. joining oppositely charged monopoles along the DW core, and the other two along $\boldsymbol{u_z}$, i. e. directed towards the surface.

The presence of $\mathbf{B^e}$ bundles allows us to propose a simplified form for eq. (2) as

$$Q = \sum_{bundles} q_n + \frac{1}{4\pi\hbar} \iint_{rest} \mathbf{B^e} \cdot d\mathbf{S} \qquad (3)$$

where the integral over a closed surface surrounding the singularity is divided into two terms from two different kinds of surface regions: the first one corresponds to a small set of bundles with a high density of emergent field flux (with typical cross-section below 70 nm × 70 nm) and the second to the rest of the surface with a much lower density of emergent field flux. Each bundle of $\mathbf{B^e}$ lines is characterized by the emergent field flux $q_n$ across its local cross-section $S_n$ as

$$q_n = \frac{1}{4\pi\hbar} \iint_{S_n} \mathbf{B^e} \cdot d\mathbf{S}_n \qquad (4)$$

With this definition, $q_n$ is equivalent to the skyrmionic charge $q_{sky}$ used for skyrmion tubes in chiral magnets [7] and for textures in thin films such as vortices, antivortices, merons or skyrmions [1,2]. The main difference is that, in the present case, the integration surface corresponds to the bundle cross-section $S_n$ rather than to the film plane. For example, the emergent field flux entering across $S_1$, $S_2$, $S_3$ and $S_4$, as indicated in Fig. 3(b), is $q_1 = -0.22$, $q_2 = -0.24$, $q_3 = -0.21$ and $q_4 = -0.16$, which accounts for more than 80% of the topological charge of Bloch point B6. Similarly, the sum of bundle charges around B7 is 94% of $Q^{B7}$ (calculated at $S_2$, $S_5$, $S_6$ and $S_7$).

An intriguing effect is the very different magnetic configuration of the emergent field bundles connecting the Bloch point dipole in spite of their very similar topological flux. Horizontal $\mathbf{B^e}$ bundles (running along $\mathbf{u}_\parallel$ within the central DW, Figs. 3(d)-(f)), consist of CW $(m_\perp, m_z)$ vortices across the thickness with a core either along $+m_\parallel$ ($S_1$ and $S_5$) or $-m_\parallel$ ($S_2$). Vertical $\mathbf{B^e}$ bundles (running along $\mathbf{u}_z$, Figs. 3(g)-(j)) correspond to regions with an approximately 90º in-plane magnetization rotation (from $m_\parallel$ to $m_\perp$) together with an out-of-plane polarity change from $+m_z$ to $-m_z$ that appear at the boundary between closure domains near the sample surface. Both vertical and horizontal bundles carry an emergent field flux $|q_n|$ close to ¼ which is half the meron charge.

We must note that magnetic vortices are usually considered as merons [2] and their skyrmionic charge is estimated as $q_{sky} = \frac{1}{2} polarity \times winding\ number = \pm\frac{1}{2}$, since polarity and winding numbers are easily inferred from qualitative contrast changes in microscopy images [2,6]. This can be a good approximation for ideal thin film nanostructures with well-defined magnetization at the boundary [1,2] but, in view of the calculated $q_n$, it fails clearly in this 3D microstructure. The key point in this reduction in topological charge is the asymmetry in the $(m_\perp, m_z)$ vortices across the thickness that form the core of the asymmetric Bloch wall. As indicated by the vertical dashed lines in Figs. 3(d)-(f), vortex cores are essentially $(m_\parallel = 1, m_\perp = 0, m_z = 0)$ lines so that they must be slightly displaced from the $m_\parallel = 0$ DW boundary (displacement of the order of 20 nm). The fact that the vortex core lies within one of the $m_\parallel$ domains, tilts part of the circulating magnetization towards the sample plane weakening the $m_z$ component at one side of the $(m_\perp, m_z)$ vortex. Therefore, the solid angle covered by the magnetization orientations within the bundle cross section becomes smaller and does not cover a hemisphere as in an ideal meron. This effect is captured by the reduced values of $|q_n|$ calculated above. It is interesting to stress that the lateral vortex core displacements are required by the condition $|\boldsymbol{m}| = 1$ except at point singularities [16], which is the basis for topological charge definition with eq. (2) in the ferromagnet. Thus, the observed $|q_1|, |q_2|, |q_5| < \frac{1}{2}$ is a direct consequence of the flux closed configuration of the planar microstructure with magnetization in the domains parallel to the sample plane and orthogonal to the flux-closure $(m_\perp, m_z)$ vortices across the thickness.

The Bloch point dipole B6-B7 corresponds to a pair of kinks in the vortex core line as it crosses from the $+m_\parallel$ to the $-m_\parallel$ domain and back. At each kink the horizontal vortex polarity is reversed, keeping a constant CW circulation, so that DW chirality is inverted. This causes a net change in the emergent field flux in the vicinity of the kink ($\Delta Q_\parallel^{B6} = q_1 + q_2 = -0.46$ upon crossing B6 and $\Delta Q_\parallel^{B7} = -q_2 + q_5 = +0.45$ upon crossing B7). The Bloch point orientation $\boldsymbol{t}$, described by the angle $\alpha$ relative to $u_\parallel$ in Table 1, is directly linked to the changes in these lateral vortex core displacements at the crossing points of $m_\parallel = 0$ and $m_z = 0$ lines: $\boldsymbol{t}$ is always tilted from the $m_\parallel = 0$ DW boundary with an alternating sign of $\alpha$ depending on polarity reversal. The "missing" emergent field flux $Q^{B6} - \Delta Q_\parallel^{B6}$, due to the reduced horizontal bundle charges, appears in the vertical bundles towards the sample surface. The asymmetric $(m_\perp, m_z)$ vortices require a Néel closure structure on top of the strongest $m_z$ branch, which changes from $-m_z$ to $+m_z$ depending on the lateral vortex core displacements. Stitching together these asymmetric closure structures, above and below the vortex line kinks, requires the half-meron textures shown in Figs. 3(g)-(j).

## C. Bloch Point Triplet and Helical Vortex.

Figure 4 shows the magnetization around the Bloch point triplet found near the microstructure apex at a transition in the central DW vortex from CCW to CW circulation (i.e., a chirality reversal without polarity change). In principle, circulation sense does not affect the vortex topological charge since a CW state can be continuously deformed into a CCW configuration [2]. However, we observe that inverting DW chirality by changing vortex circulation sense implies a more complex reorganization of the magnetization in the microstructure than the chirality transitions by polarity reversals previously discussed. The $+m_z$ (dark contrast) and $-m_z$ (bright contrast) branches of the $(m_\perp, m_z)$ vortices across the thickness exchange places with a lateral zig - zag in order to switch from CCW to CW circulation. This zig-zag is observed directly in the MTXM projection image at normal incidence (Fig. 3(b)), and also in the $m_z$ map of the central $z = 0$ plane (Fig. 4(c)).

The CCW-CW transition is mediated by B3-B4, two Bloch points with topological charges of the same sign (see Table 1) and only 80 nm apart. This implies a very strong repulsion of emergent field lines in the region between them, as shown in Fig. 4(d). The emergent field flux concentrates in bundles with a preferential vertical orientation. At the top sample surface ($z = 70$ nm plane, Fig. 4(e)), two intense green spots with topological flux $q_{10} = -0.40$ and $q_{11} = -0.28$ mark the emergence of $\mathbf{B^e}$ vertical bundles coming from B3 and B4 at the $+m_z/-m_z$ boundaries along $\mathbf{u}_\perp$. Similarly, just below B3-B4 ($z = 0$ nm plane, Fig. 4(f)), there are two purple spots with $q_{13} = -0.17$ and $q_{14} = -0.33$ (sign considers an outward surface element, i.e. pointing along $-z$). These also correspond to emergent field bundles coming from B3 and B4 that define the boundaries of the small $-m_z$ transverse region between the two large $\pm m_\parallel$ domains. A closer look at the 3D magnetization around the emergent field bundle $q_{14}$ shows a low symmetry helical vortex that combines $\pm m_\parallel$, $\pm m_\perp$ and $\pm m_z$ rotations at an oblique angle relative to the sample surface (Fig. 4 (g)), so that it cannot be easily classified in the usual schemes in thin films of vortex/antivortex textures with a well-defined polarity and circulation sense. Part of the $\boldsymbol{B}_e$ lines in $q_{14}$ connect with the positively charged B5, where the horizontal bundle corresponding to the CW vortex in the central domain wall departs towards the right, while others intersect with the bottom sample surface to account for the magnetization textures of the flux closure domain structure.

Topological charge conservation, as stated in eq. (2), implies that emergent field lines must terminate either at a Bloch point or at the sample surface. As a consequence, in an infinite sample (without surfaces) a CW-CCW transition could occur either within a continuous $\boldsymbol{B}_e$ bundle via unwinding and rewinding the vortex circulation through an intermediate radial vortex state, or via a topological dipole combining a radial hedgehog and a circulating Bloch point joined by a $\boldsymbol{B}_e$ bundle. However, our results show that these simple solutions are avoided in this real permalloy microstructure, since they would include radial textures across the sample thickness with a very high cost in magnetostatic energy. The system prefers instead to break the emergent field bundles, directing them towards the surface using the topological repulsion between two topological charges of the same sign.

**IV. CONCLUSSIONS**

The quantitative information contained in the experimental $m(r)$ reconstructed by XVMT can significantly improve our understanding of 3D magnetic systems. Here, different magnetic textures such as Bloch point dipoles and a triplet, asymmetric and helical vortices, half merons, etc. have been observed at the chirality transitions of a 3D DW within a low symmetry permalloy microstructure. At these real magnetic singularities shape effects severely modify the ideal spherical symmetry of isolated monopoles and create bundles of emergent field lines that either connect topological dipoles (as in B1-B2 or B6-B7 dipoles) or tend towards the surface (as in the pair of negatively charged monopoles B3-B4). These bundles are associated with 3D low symmetry magnetic textures carrying a fractional topological flux $q_n$, and provide a simple description of topological constraints derived from total charge conservation in complex magnetic configurations.

**ACKNOWLEDGEMENTS**

The ALBA Synchrotron is funded by the Ministry of Research and Innovation of Spain, by the Generalitat de Catalunya and by European FEDER funds. This project has been supported by Spanish MICIN under grant PID2019-104604RB/AEI/10.13039/501100011033 and by Asturias FICYT under grant AYUD/2021/51185 with the support of FEDER funds.

**TABLES AND FIGURES**

| Bloch point | Z (nm) | Polarity reversal | Core orientation | | Circulation sense | Q |
| :---: | :---: | :---: | :---: | :---: | :---: | :---: |
| | | | α (deg) | γ (deg) | | |
| B1 | 20 | H2H | 54 | -6 | CW | -0,93 |
| B2 | 30 | T2T | -42 | -6 | CW | 0,95 |
| B3 | 40 | H2H | 69 | 6 | CCW | -1,01 |
| B4 | 50 | H2H | 72 | -46 | CW | -0,9 |
| B5 | 20 | T2T | -46 | -9 | CW | 0,99 |
| B6 | 0 | H2H | 15 | -18 | CW | -0,96 |
| B7 | -10 | T2T | -25 | -14 | CW | 1,03 |

**Table 1.** Main characteristics of representative Bloch points within central DW: Vertical coordinate $z$ relative to sample centre (with 10 nm accuracy given by pixel size in the reconstructed volume), polarity reversal (H2H or T2T), orientation of core vector $t$ relative to $u_\parallel, u_\perp, u_z$ with the angles $\alpha$ and $\gamma$ as defined in Fig. 2, circulation sense (CW or CCW), and topological charge calculated from eq. (2) as the emergent field flux across a ($L \times L \times L$) closed box with lateral size $L$= 70 nm. $Q(B4)$ is calculated as $Q(B4) = Q(B4 + B3) - Q(B3)$ using a larger box containing both B3 and B4 since the latter is too close to the top surface of the microstructure and to B3 to resolve its topological charge individually. Note the large spread in the values of $\alpha$ (the in-plane angle between $t$ and $u_\parallel$), indicating the tendency of $t$ to follow the transverse segments of kinks in the $m_z = 0$ boundary, and the clear correlation between the sign of $\alpha$ and polarity reversal at the Bloch point: H2H for positive $\alpha$ and T2T for negative $\alpha$.

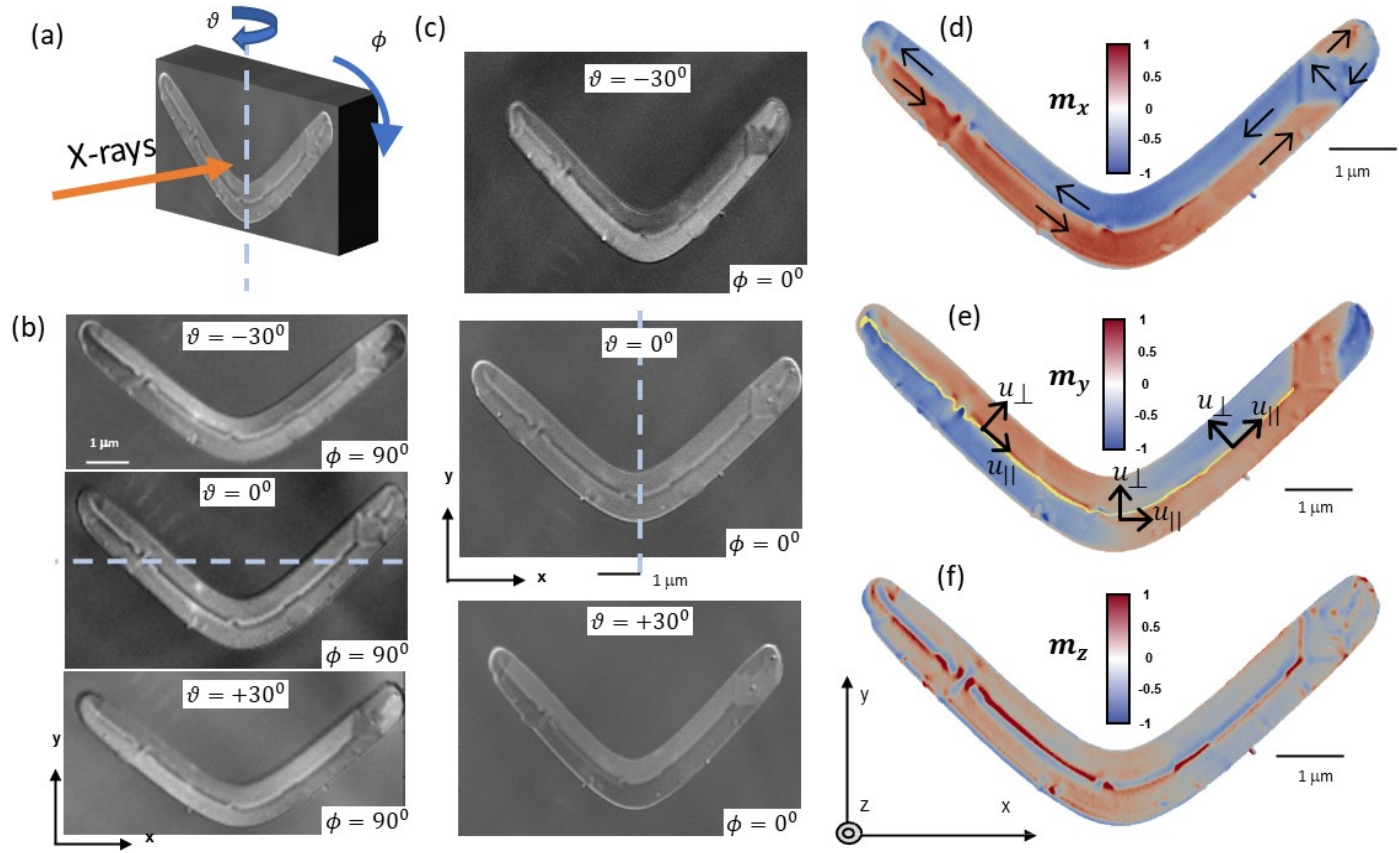

**Figure 1.** Magnetic configuration of 140 nm thick permalloy microstructure by X-ray vector tomography: a) Sketch of sample configuration at MISTRAL microscope; (b-c) MTXM projection images with different angles of incidence: (b) $\phi = 90^0$ and predominant $m_y - m_z$ contrast and (c) $\phi = 0^0$ and predominant $m_x - m_z$ contrast. Dashed lines indicate rotation axes. Top view of tomographic reconstruction of magnetic configuration at the middle plane of the sample ($z = 0$) with d) $m_x$ contrast, e) $m_y$ contrast and f) $m_z$ contrast. Arrows in (d) indicate average magnetization orientation at each domain. Local reference system $(u_\parallel, u_\perp)$ is sketched in (e) together with a solid yellow line indicating the position of the $m_\parallel = 0$ boundary that defines the central DW.

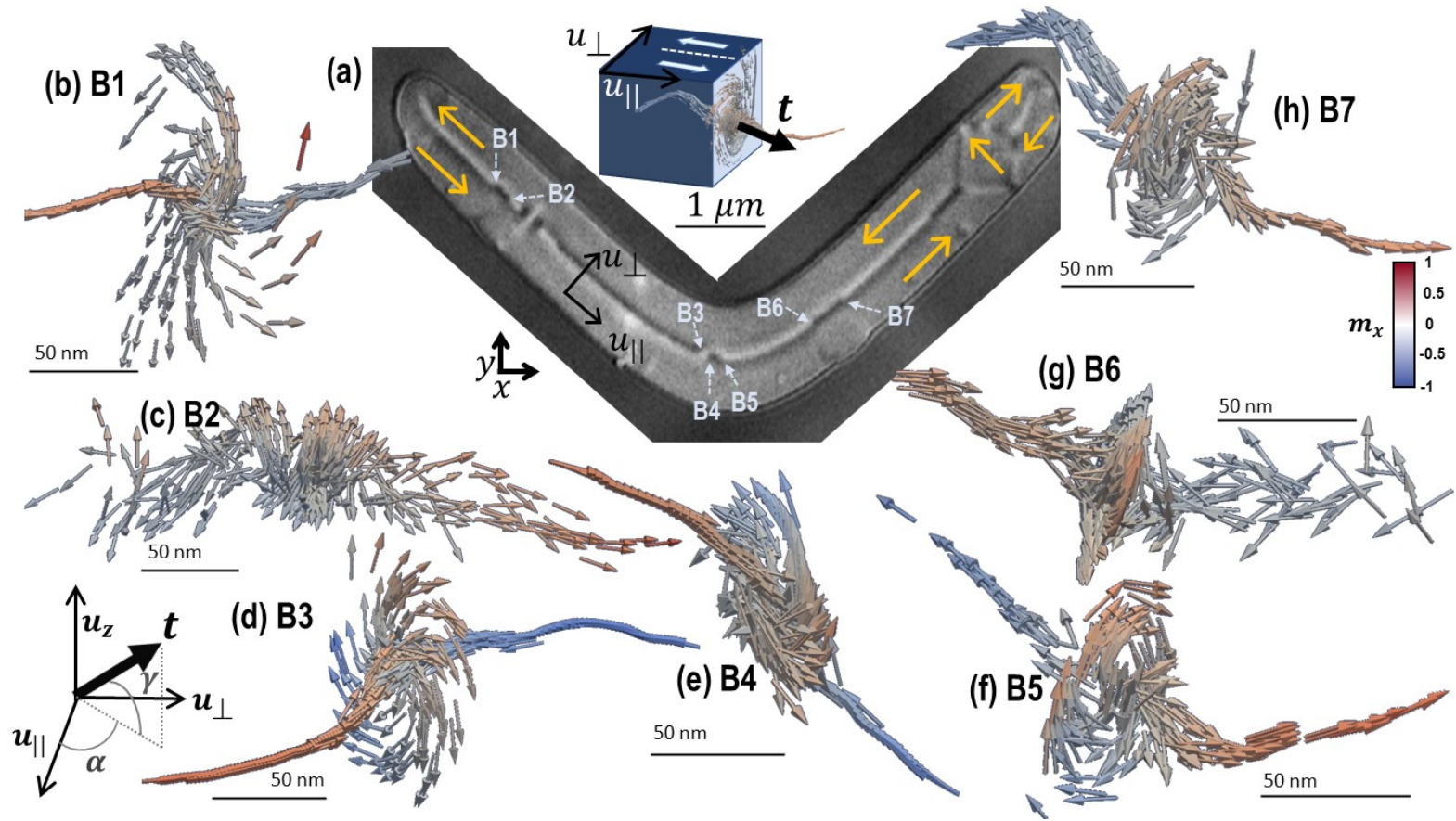

**Figure 2.** Bloch points along central DW: a) MTXM image at normal incidence sensitive only to out-of-plane magnetization component $m_z$. Dashed arrows indicate locations of selected Bloch points. Yellow arrows indicate average in-plane magnetization orientation at each domain. (b-h) Detail of the 3D measured magnetisation vector $m(r)$ around Bloch points B1-B7 showing their circulating configuration. Arrows colour corresponds to $m_x$ component. Insets show sketches of DW local reference system $u_\parallel, u_\perp, u_z$ and Bloch point core orientation vector $t$ (defined to be parallel to the central H2T or T2T line with a positive projection along $u_\parallel$, as illustrated in the sketches).

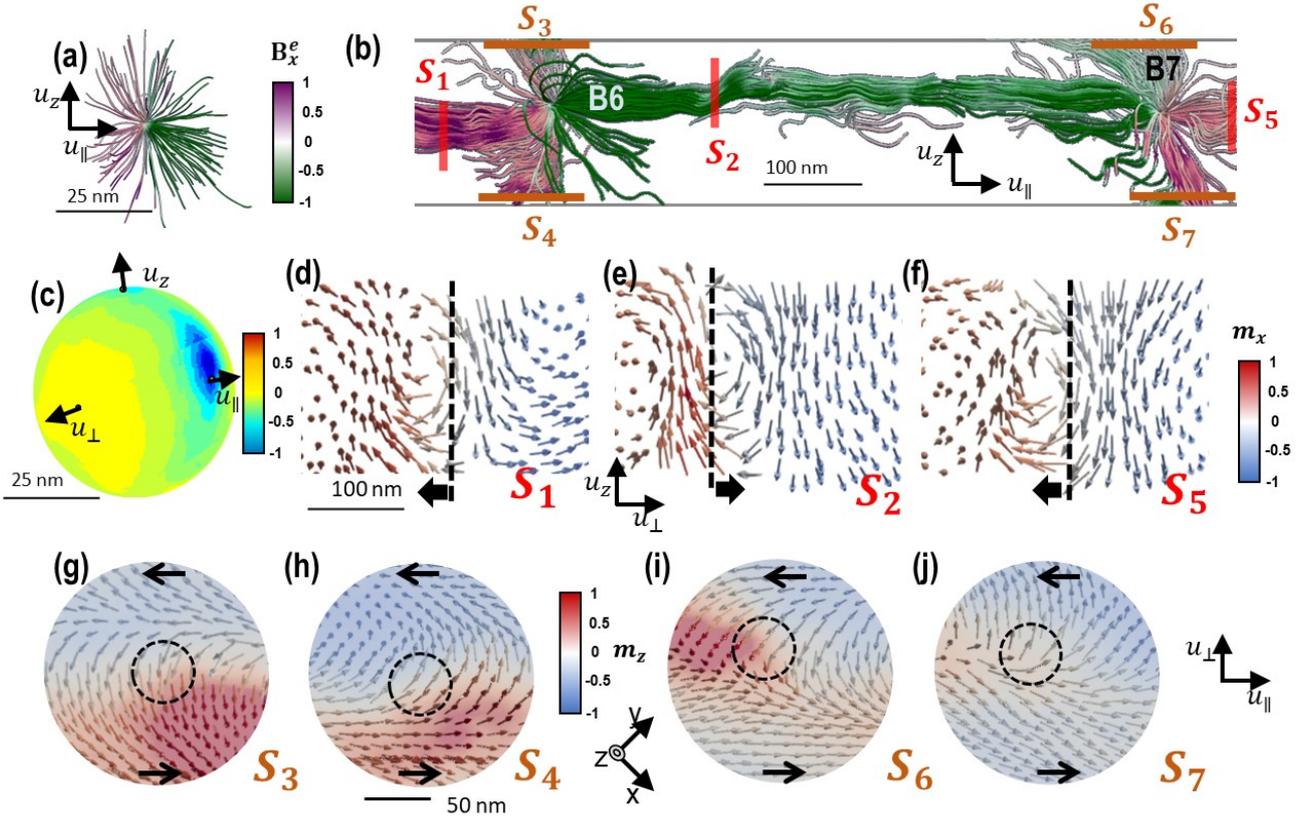

**Figure 3:** Topological dipole and emergent field bundles of fractional $q_n$ at a DW chirality transition by polarity reversal: (a) Detail of emergent field lines at B6. (b) Overview of emergent field lines connecting B6 and B7 along a longitudinal cross section of the central DW. Note the additional bundles of $\mathbf{B^e}$ lines emerging towards the sample surface at the Bloch points. Solid lines indicate the location of $S_1$ to $S_7$, the 70 $nm \times$ 70 $nm$ areas used to evaluate emergent field flux. (c) Density of emergent field flux $\mathbf{B^e} \cdot d\mathbf{S}$ (in arbitrary units) at a 50 nm diameter sphere centred in B6. (d-f) Detail of magnetic configuration of central DW at $S_1$, $S_2$ and $S_5$. Dashed lines indicate DW centre ($m_{\parallel} = 0$) at $z = 0$ and black arrows the displacement of vortex core respect to it. Note the asymmetry between $\pm m_z$ branches of the vortices across the thickness. This reduces the total angle covered in the Bloch sphere and, thus, the topological flux in the bundle to well below the ½ value of an ideal vortex. (g-j) Detail of magnetic configuration at $S_3$, $S_4$, $S_6$ and $S_7$. Circles mark the location of emergent field bundles with $q_n \approx \frac{1}{4}$ corresponding to a 90º rotation of the in-plane magnetization simultaneous to an out-of-plane polarity change.

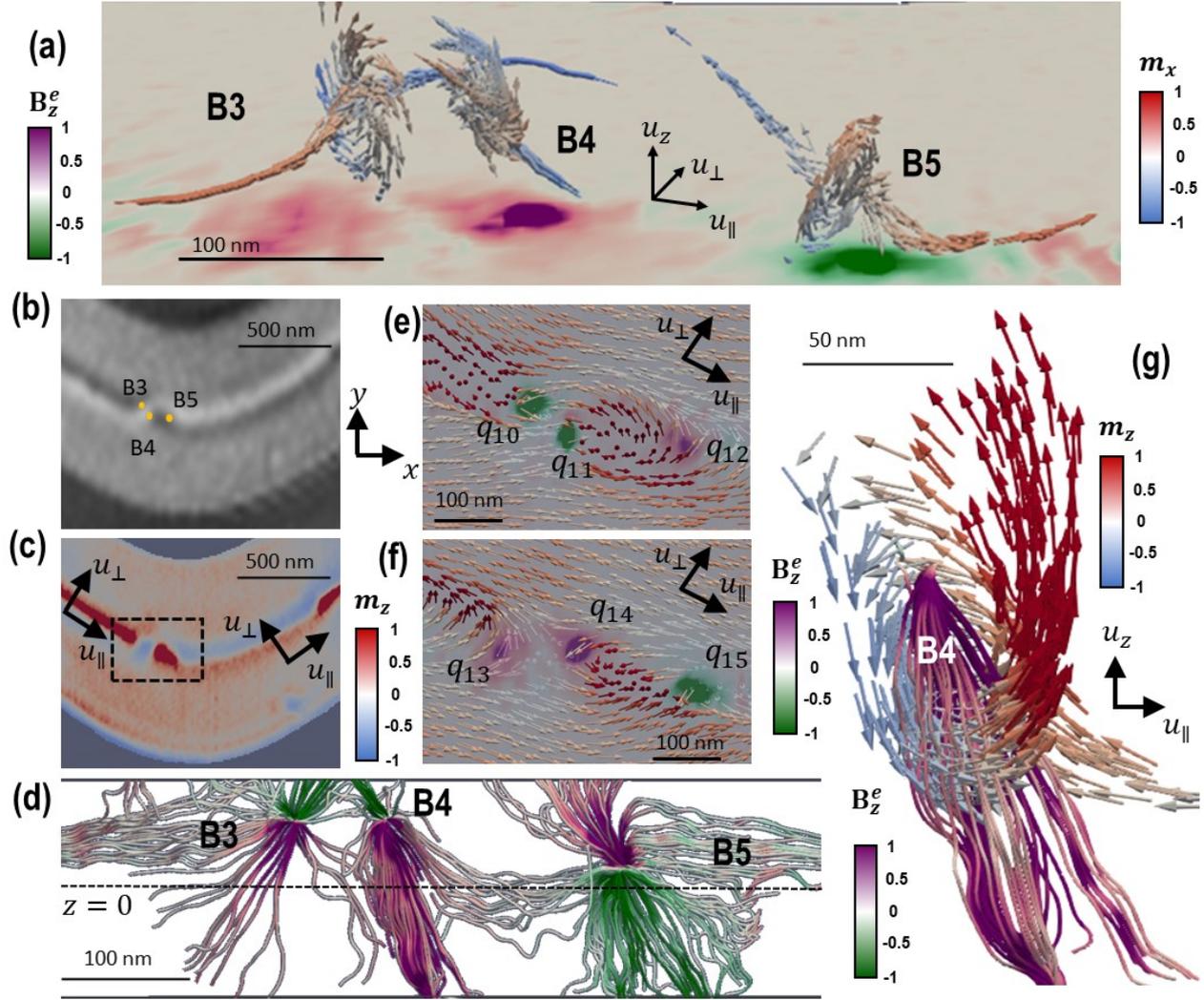

**Figure 4:** Bloch point triplet for DW chirality transformation by CW-CCW reversal: (a) Magnetic configuration of a Bloch point triplet composed of two negative monopoles (B3-B4) and a positive one (B5) at a DW transition from CCW to CW circulation sense. (b) MTXM projection image at normal incidence showing spatial location of B3, B4 and B5. (c) Top view $m_z$ map at central $z = 0$ plane. Note the zig-zag configuration of the $-m_z$ blue domain that switches the circulation sense of the central DW from CCW to CW. (d) Emergent field lines around B3, B4 and B5. Note the strong repulsion of $\mathbf{B^e}$ lines in the region between the negatively charged B3-B4 that directs the bundles towards the top/bottom surfaces of the microstructure. Zoom view of magnetic configuration at the dashed box in (c): (e) $z = 70$ nm just above the Bloch point triplet and (f) $z = 0$ nm, just below it. Green/purple shades indicate the regions of high $B_z^e$, where vertical bundles cross the $z = constant$ planes. (g) Helical vortex associated with the emergent field bundle with $q_{14} = -0.33$ below B4.